\documentstyle[12pt]{article}

\begin{document}
\begin{center}
{\LARGE{\bf Quantum Properties of General Gauge Theories\\
            with Composite and External Fields}}\\
\vspace{0.7 cm}
{\large S.~Falkenberg, B. Geyer}\footnote{E-mail:
geyer@rz.uni-leipzig.de}\\
{\small{\it Center of Theoretical Sciences, and Institute
of Theoretical Physics, \\Leipzig University,
Augustusplatz 10--11, D-04109 Leipzig, Germany}}\\
and\\
{\large P. Moshin}\\
{\small
{\it Tomsk State Pedagogical University, Tomsk 634041, Russia}}\\
\end{center}
\vspace{0.1 cm}
\begin{quotation}
 \small\noindent
 The generating functionals of Green's functions with composite and external
 fields are considered in the framework of BV and BLT quantization methods
 for general gauge theories. The corresponding Ward identities are derived
 and the gauge dependence is investigated.
\end{quotation}

\section{Introduction}
 The most general rules for manifestly covariant quantization of gauge
 theories in the path integral approach are provided by the BV formalism
 \cite{BV}, based on the principle of BRST symmetry \cite{BRST}, as well
 as by its Sp(2)-covariant version, the BLT quantization scheme \cite{BLT},
 based on the principle of extended BRST symmetry \cite{extBRST}. These
 methods currently underly the study of quantum properties of arbitrary
 (general) gauge theories in the Lagrangian formalism, either immediately
 providing the corresponding basis (e.g. derivation of the Ward identities,
 study of renormalization and gauge dependence \cite{LVT}, analysis of
 unitarity conditions \cite{LM}) or playing a key role in the
 interpretation of alternative quantization methods (e.g. triplectic
 \cite{BMS}, superfield \cite{LMR}, osp(1,2)-covariant quantization
 \cite{GLM}).
 In particular, the methods \cite{BV,BLT} have been used to analyse the
 quantum structure of general gauge theories with composite fields
 \cite{LO,LOR} (in the BV and BLT formalisms, respectively); for these
 theories the corresponding Ward identities were derived and the related
 issue of gauge dependence was investigated. Also based on the use of the
 quantization methods \cite{BV,BLT} was the recent study of Ref.~\cite{ext},Πwhich carried out an investigation of the Ward identities and gauge
 dependence for general gauge theories with external fields.

 In particular, the studies of Refs.~\cite{LOR,ext} revealed the fact that
 the gauge dependence for theories with composite or external fields
 is described with the help of certain fermionic operators in the BV
 formalism, and doublets of fermionic operators in the BLT formalism.
 At the same time, a remarkable feature, commonly shared by both types
 of theories, consists in the property of (generalized) nilpotency of the
 fermionic operators in question.

 The purpose of this paper is to provide an extension of the studies
 of Refs.~\cite{LO,LOR,ext}, which is aimed at incorporating composite
 fields, simultaneously with external ones, into arbitrary quantum gauge
 theories. The reason to consider this very general setting consists in
 the following.
 The procedure for constructing the effective action with composite fields
 \cite{comp} (see also Ref.~\cite{ref}) offers a wide range of applications
 to quantum field theory models. Among the most important of them we find the
 study of such phenomenologically relevant theories as the Standard Model
 \cite{StMod} and models of the inflationary Universe \cite{infUniv}, as well
 as SUSY theories \cite{SUSY} and theory of strings \cite{string}. Recent
 activities in supersymmetric YM theories and other gauge models
 have signalled the relevance of their extension to the case of composite
 fields on external background. The external field approach is applied
 mainly to handle certain difficulties in the path integral formulation
 of quantum theories by means of lifting the functional integration from
 a part of the variables, which are afterwards considered as external
 parameters. In the absence of a consistent theory of quantum gravity, this
 approach currently has the status of an indispensable tool providing
 insight into numerous problems which arise in the physics of black holes,
 as well as permitting to incorporate gravitational effects into the
 cosmology of the early Universe.

 In this paper, the generating functionals of Green's functions with
 composite and external fields combined are considered in the framework
 of the BV (section 2) and BLT (section 3) quantization methods for general
 gauge theories. For these functionals we derive the corresponding Ward
 identities and investigate the most general form of gauge dependence.
 It is revealed that combining composite and external fields into one
 scheme gives rise to a radical change in the character of both the Ward
 identities and gauge dependence (as compared to those obtained in
 Refs.~\cite{LO,LOR,ext}). In particular, it is shown that in the most
 general case of a quantum theory with composite and external fields the
 operators describing the gauge dependence suffer from the violation of
 (generalized) nilpotency.

 We use De Witt's condensed notations \cite{DeWitt} as well as notations
 adopted in Refs.~\cite{BV,BLT}. The invariant tensor of the group Sp(2),
 which is a constant antisymmetric second-rank tensor, is denoted as
 $\varepsilon^{ab}$ ($a=1, 2$), with the normalization $\varepsilon^{12}=1$.ΠSymmetrization over Sp(2) indices is denoted as
 $A^{\{ab\}}=A^{ab}+A^{ba}$. Derivatives with respect to sources and
 antifields are understood as acting from the left, and those to fields,
 as acting from the right (unless otherwise specified); left-hand
 derivatives with respect to the fields are labelled by the subscript
 ``${\it l}\,$" ($\delta_l/\delta\phi$ stands for the left-hand
 derivative with respect to the field $\phi$).


\section{Quantum Gauge Theories with Composite and External Fields in the BV
        Formalism}
 We first recall that the quantization of a gauge theory within the BV
 approach \cite{BV} requires introducing a complete set of fields $\phi^A$
 and a set of corresponding antifields $\phi^*_A$ (which play the role
 of sources of BRST transformations), with Grassmann parities
\[
 \varepsilon(\phi^A)\equiv\varepsilon_A,\;\;
 \varepsilon(\phi^*_A)=\varepsilon_A+1.
\]

 The content of the configuration space of the fields $\phi^A$ (composed by
 the initial classical fields, the (anti) ghost pyramids and the Lagrangian
 multipliers) is determined by the properties of the original classical
 theory, i.e. by the linear dependence (for reducible theories) or
 independence (for irreducible theories) of the generators of gauge
 transformations.

 In terms of the variables $\phi^A$ we define composite fields
 $\sigma^m(\phi)$, i.e.
\begin{eqnarray*}
 \sigma^m(\phi)=
 \sum_{k \geq 2}\frac{1}{k!}\Lambda^m_{A_1...A_k}\phi^{A_1}...\phi^{A_k},
 \;\;\varepsilon(\sigma^m)\equiv\varepsilon_m,
\end{eqnarray*}
 where $\Lambda^m_{A_1...A_k}$ are some field-independent coefficients.

 The extended generating functional $Z(J,L,\phi^*)$ of Green's functions with
 composite fields is constructed within the BV quantization approach by the
 rule (see, for example, Ref.~\cite{LO})
\begin{eqnarray}
 Z(J,L,\phi^*)&=&\int d\phi\;\exp\bigg\{\frac{i}{\hbar}\bigg(
 S_{\rm ext}(\phi,\phi^{*})+
 J_A\phi^A+L_m\sigma^m(\phi)\bigg)\bigg\},
\end{eqnarray}
 where $J_A$ are the usual sources of the
 fields $\phi^A$, $\varepsilon(J_A)=\varepsilon_A$;
 $L_m$ are the sources of the composite fields $\sigma^m(\phi)$,
 $\varepsilon(L_m)=\varepsilon_m$; and $S_{\rm ext}=
 S_{\rm ext}(\phi,\phi^{*})$Πis the gauge-fixed quantum action defined in the usual manner as
\begin{equation}
 \exp\bigg\{\frac{i}{\hbar}S_{\rm ext}\bigg\}=\exp\bigg(\hat{T}(\Psi)\bigg)
 \exp\bigg\{\frac{i}{\hbar}S\bigg\}.
\end{equation}

 In eq. (2), $S=S(\phi,\phi^*)$ is a bosonic functional satisfying the
 equation
\begin{equation}
 \frac{1}{2}(S,S)=i\hbar\Delta S,
\end{equation}
 or equivalently
\begin{equation}
 \Delta\exp\bigg\{\frac{i}{\hbar}S\bigg\}=0,
\end{equation}
 with the boundary condition
\begin{eqnarray*}
 S|_{\phi^{*}=\hbar=0}={\cal S},
\end{eqnarray*}
 where $\cal S$ is the original gauge-invariant classical action.
 At the same time, the operator $\hat{T}(\Psi)$ has the form
\begin{equation}
 \hat{T}(\Psi)=[\Delta,\Psi]_{+}\;,
\end{equation}
 where $\Psi$ is a fermionic gauge-fixing functional.

 In eqs.~(3)--(5) we use the standard definition of the antibracket, given
 for two arbitrary functionals $F=F(\phi,\phi^*)$, $G=G(\phi,\phi^*)$ by the
 rule
\begin{eqnarray*}
 (F,\;G)=\frac{\delta F}{\delta\phi^A}\frac{\delta G}{\delta
 \phi^*_A}-(-1)^{(\varepsilon(F)+1)(\varepsilon(G)+1)}
 \frac{\delta G}{\delta\phi^A}\frac{\delta F}
 {\delta\phi^*_A}\,,
\end{eqnarray*}
 as well as the usual definition of the operator $\Delta$
\begin{eqnarray*}
 \Delta=(-1)^{\varepsilon_A}\frac{\delta_l}{\delta\phi^A}\frac
 {\delta}{\delta\phi^{*}_A}\,,
\end{eqnarray*}
 possessing the property of nilpotency $\Delta^2=0$.

 It is well-known that the
 gauge-fixing (2), (5) represents a particular case of transformation
 corresponding to any fermionic operator (chosen for $\Psi$) and
 describing the arbitrariness in solutions of eq.~(3) (or eq.~(4)), namely,
\begin{eqnarray}
 {\Delta}\exp\bigg\{\frac{i}{\hbar}S_{\rm ext}\bigg\}=0.
\end{eqnarray}
 In what follows we consider the most general case of gauge-fixing,
 corresponding to an arbitrary operator-valued fermionic functional $\Psi$Π(clearly, it should be chosen in such a way as to ensure the existence
 of the functional integral).

 Let us now consider the following representation of the generating
 functional $Z(J,L,\phi^*)$ in eq.~(1):
\begin{eqnarray*}
 Z(J,L,\phi^*)=\int d\psi\;{\cal Z}({\cal J},L,\psi,\phi^*)\exp\bigg
 (\frac{i}{\hbar}{\cal Y}\psi\bigg),
\end{eqnarray*}
 where
\begin{eqnarray}
 {\cal Z}({\cal J},L,\psi,\phi^*)=\int d\varphi\;\exp\bigg\{\frac{i}
 {\hbar}\bigg(S_{\rm ext}(\varphi,\psi,\phi^{*})+{\cal J}\varphi
 +L\sigma(\varphi,\psi)\bigg)
 \bigg\}.
\end{eqnarray}
 Given this, we have assumed the decomposition
\[
 \phi^A=(\varphi^i,\;\psi^\alpha),\;\;\;
 J_A=({\cal J}_i,\;{\cal Y}_\alpha),\\
\]
\[
 \varepsilon(\varphi^i)\equiv\varepsilon_i,\;\;\;
 \varepsilon(\psi^\alpha)\equiv\varepsilon_\alpha.
\]
 In what follows we refer to ${\cal Z}={\cal Z}({\cal J},L,\psi,\phi^*)$
 as the extended generating functional of Green's functions with composite
 fields on the background of external fields $\psi^\alpha$.
 Clearly, the validity of the functional integral in eq.~(1) implies the
 existence of the integral in eq.~(7) without any restriction on the
 structure of the subspace $\psi^\alpha$.
 At the same time, the composite fields $\sigma^m (\varphi,\psi)$,
 according to their original definition, may be considered as given by
\begin{eqnarray*}
 \sigma^m(\varphi, \psi)=
 \sum_{k \geq 2}\frac{1}{k!}\tilde \Lambda^m_{i_1\ldots i_k} (\psi)
 \varphi^{i_1} \ldots \varphi^{i_k},
\end{eqnarray*}
 with the arising coefficients now depending on the external fields
 $\psi^\alpha$. This, as will be shown more explicitly below, leads to additional
 peculiarities, which are absent in the studies of Ref.~\cite{LO,LOR,ext}.

 The Ward identities for a general gauge theory with composite and external
 fields considered in the BV quantization scheme can be obtained as direct
 consequence of equation (6) for the gauge-fixed quantum action
 $S_{\rm ext}$. Namely, integrating eq.~(6) over the fields $\varphi^i$ with
 the weight functional
\begin{eqnarray*}
 \exp\bigg\{\frac{i}{\hbar}\bigg[{\cal J}_i\varphi^i+L_m
 \sigma^m(\varphi,\psi)
 \bigg]\bigg\},Œ\end{eqnarray*}
 we have
\begin{eqnarray}
 \int d\varphi\;\exp\bigg[
 \frac{i}{\hbar}\bigg({\cal J}_i\varphi^i+
 L_m\sigma^m(\varphi,\psi)
 \bigg)
 \bigg]
 \Delta\exp\bigg\{\frac{i}{\hbar}S_{\rm ext}(\varphi,\psi,\phi^{*})
 \bigg\}=0.
\end{eqnarray}
 Next, performing in eq.~(8) integration by parts, with allowance for the
 relation
\[
 \exp\bigg\{\frac{i}{\hbar}\bigg(
 {\cal J}_i\varphi^i+L_m\sigma^m(\varphi,\psi)
 \bigg)\bigg\}
 \Delta=
\]
\begin{eqnarray}
 =\bigg(\Delta-\frac{i}{\hbar}{\cal J}_i
 \frac{\delta}{\delta\varphi^{*}_i}-
 \frac{i}{\hbar}L_m\sigma^m_{,A}(\varphi,\psi)
 \frac{\delta}{\delta\phi^{*}_A}\bigg)
 \exp\bigg\{\frac{i}{\hbar}\bigg({\cal J}_i\varphi^i+
 L_m\sigma^m(\varphi,\psi)\bigg)\bigg\},
\end{eqnarray}
\[
 \sigma^m_{,A}(\varphi,\psi)\equiv\frac{\delta}{\delta\phi^{A}}
 \sigma^m(\varphi,\psi),
\]
 we arrive at the following Ward identities for
 ${\cal Z}={\cal Z}({\cal J},L,\psi,\phi^*)$:
\begin{eqnarray}
 \hat{\omega}{\cal Z}=0,
\end{eqnarray}
 where $\hat{\omega}$ stands for the operator
\begin{eqnarray}
 \hat{\omega}=i\hbar\Delta_\psi+{\cal J}_i\frac{\delta}
 {\delta\varphi^{*}_i}+L_m\sigma^m_{,A}\bigg(\frac{\hbar}{i}
 \frac{\delta}{\delta{\cal J}},\psi\bigg)
 \frac{\delta}{\delta\phi^{*}_A}\,,\;\;\;
 \Delta_\psi\equiv(-1)^{\varepsilon_\alpha}\frac{\delta_l}{\delta\psi^
 \alpha}\frac{\delta}{\delta\psi^{*}_\alpha}\,.
\end{eqnarray}

 At the same time, the fact that the composite fields
 $\sigma^m(\varphi,\psi)$ now depend on the external fields $\psi^\alpha$
 implies that the nilpotency of $\hat{\omega}$ becomes in general
 violated, i.e.
\[Π\hat{\omega}^2=i\hbar(-1)^{\varepsilon_i}L_m
 \sigma^m_{,i\alpha}\left(\frac{\hbar}{i}
 \frac{\delta}{\delta{\cal J}},\psi\right)
 \frac{\delta}{\delta\psi^*_\alpha}\frac{\delta}{\delta\varphi^*_i}\;,
\]
 where
\[
 \sigma^m_{,i\alpha}\left(\frac{\hbar}{i}
 \frac{\delta}{\delta{\cal J}},\psi\right)\equiv
 \left.
 \frac{\delta}{\delta\psi^\alpha}\sigma^m_{,i}(\varphi,\psi)
 \right|_{ \varphi=\frac{\hbar}{i}\frac{\delta}{\delta{\cal J}}}\;.
\]

 In terms of the generating functional
 ${\cal W}={\cal W}({\cal J},L,\psi,\phi^*)$,
\[
 {\cal Z}=\exp\bigg\{\frac{i}{\hbar}{\cal W}\bigg\},
\]
 of connected Green's functions with composite and external fields,
 the identities~(10) take on the form
\begin{eqnarray}
 \hat{\Omega}{\cal W}=\frac{\delta{\cal W}}{\delta\psi^\alpha}
 \frac{\delta{\cal W}}{\delta\psi^*_\alpha}\,,
\end{eqnarray}
\[
 \hat{\Omega}=
 i\hbar\Delta_\psi+{\cal J}_i\frac{\delta}
 {\delta\varphi^{*}_i}+L_m\sigma^m_{,A}\bigg(\frac{\delta{\cal W}}
 {\delta{\cal J}}
 +\frac{\hbar}{i}\frac{\delta}{\delta{\cal J}},\psi\bigg)
 \frac{\delta}{\delta\phi^{*}_A}.
\]
 In order to introduce the (extended) generating functional of 1PI vertex
 functions with composite and external fields we make use of the
 standard definition \cite{comp} of the generating functional of vertex
 functions with composite fields, which admits of a natural generalization
 to the case of external fields. Namely, let us introduce the generating
 functional in question by means of the following Legendre transformation
 with respect to the sources ${\cal J}_i$, $L_m$:
\[
 \Gamma(\varphi,\Sigma,\psi,\phi^*)={\cal W}({\cal J},L,\psi,\phi^*)-
 {\cal J}_i\varphi^i-L_m\bigg(\Sigma^m+\sigma^m(\varphi,\psi)\bigg),
\]
 where
\[
 \varphi^i=\frac{\delta{\cal W}}{\delta{\cal J}_i},\;\;\;
 \Sigma^m=\frac{\delta{\cal W}}{\delta L_m}
 -\sigma^m\bigg(\frac{\delta{\cal W}}{\delta{\cal J}},\psi\bigg).
\]
 Given this, we haveŒ\[
 {\cal J}_i=-\frac{\delta\Gamma}{\delta\varphi^i}+
 \frac{\delta\Gamma}{\delta\Sigma^m}\sigma^m_{,i}(\varphi,\psi),\;\;\;
 L_m=-\frac{\delta\Gamma}{\delta\Sigma^m}\,.
\]
 From the definition of the Legendre transformation it follows that
\begin{eqnarray*}
 \left.\frac{\delta}{\delta{\cal J}_i}\right|_{L,\psi,\phi^*}&=&
 \left.\frac{\delta\varphi^j}{\delta{\cal J}_i}
 \frac{\delta_l}{\delta\varphi^j}\right|_{\Sigma,\psi,\phi^*}
 +
 \left.\frac{\delta\Sigma^m}{\delta{\cal J}_i}
 \frac{\delta_l}{\delta\Sigma^m}\right|_{\varphi,\psi,\phi^*},\\
 \left.\frac{\delta}{\delta L_m}\right|_{{\cal J},\psi,\phi^*}&=&
 \left.\frac{\delta\varphi^i}{\delta L_m}
 \frac{\delta_l}{\delta\varphi^i}\right|_{\Sigma,\psi,\phi^*}
 +
 \left.\frac{\delta\Sigma^m}{\delta L_m}
 \frac{\delta_l}{\delta\Sigma^m}\right|_{\varphi,\psi,\phi^*},\\
 \left.\frac{\delta_l}{\delta\psi^\alpha}\right|_{{\cal J},L,\phi^*}&=&
 \left.\frac{\delta_l}{\delta\psi^\alpha}\right|_{\varphi,\Sigma,\phi^*}+
 \left.\frac{\delta_l\varphi^i}{\delta\psi^\alpha}
 \frac{\delta_l}{\delta\varphi^i}\right|_{\Sigma,\psi,\phi^*}
 +
 \left.\frac{\delta_l\Sigma^m}{\delta\psi^\alpha}
 \frac{\delta_l}{\delta\Sigma^m}\right|_{\varphi,\psi,\phi^*},\\
 \left.\frac{\delta}{\delta\phi^*_A}\right|_{{\cal J},L,\psi}&=&
 \left.\frac{\delta}{\delta\phi^*_A}\right|_{\varphi,\Sigma,\psi}+
 \left.\frac{\delta\varphi^i}{\delta\phi^*_A}
 \frac{\delta_l}{\delta\varphi^i}\right|_{\Sigma,\psi,\phi^*}
 +
 \left.\frac{\delta\Sigma^m}{\delta\phi^*_A}
 \frac{\delta_l}{\delta\Sigma^m}\right|_{\varphi,\psi,\phi^*}.
\end{eqnarray*}
 Then, by virtue of eq.~(12) and the relations
\[
 \frac{\delta{\cal W}}{\delta\psi^\alpha}=
 \frac{\delta\Gamma}{\delta\psi^\alpha}
 -
 \frac{\delta\Gamma}{\delta\Sigma^m}\sigma^m_{,\alpha}(\varphi,\psi),\;\;\;
 \frac{\delta{\cal W}}{\delta\phi^*_A}=
 \frac{\delta\Gamma}{\delta\phi^*_A}\,,
\]
 we arrive at the following Ward identities for the functional
 $\Gamma(\varphi,\Sigma,\psi,\phi^*)$:
\begin{eqnarray}
 &&
 \frac{1}{2}(\Gamma,\Gamma)+\frac{\delta\Gamma}{\delta\Sigma^m}Π\bigg(\sigma^m_{,A}(\hat{\varphi},\psi)
 -\sigma^m_{,A}(\varphi,\psi)\bigg)\frac{\delta\Gamma}{\delta\phi^*_A}
 =\nonumber\\
 && =
 i\hbar\bigg\{
 \Delta_\psi\Gamma-
 (-1)^{\varepsilon_\alpha\varepsilon_\rho}
 (G^{''-1})^{\rho\sigma}
 \bigg[
 \frac{\delta_l}{\delta\Phi^\sigma}
 \bigg(
 \frac{\delta\Gamma}{\delta\psi^\alpha}
 -
 \frac{\delta\Gamma}{\delta\Sigma^m}\sigma^m_{,\alpha}(\varphi,\psi)
 \bigg)
 \bigg]
 \frac{\delta_l}{\delta\Phi^\rho}
 \frac{\delta\Gamma}{\delta\psi^*_\alpha}\nonumber\\
 &&
 +
 (-1)^{\varepsilon_i(\varepsilon_\alpha+\varepsilon_m+1)}
 (G^{''-1})^{i\sigma}
 \bigg[
 \frac{\delta_l}{\delta\Phi^\sigma}
 \bigg(
 \frac{\delta\Gamma}{\delta\psi^\alpha}
 -
 \frac{\delta\Gamma}{\delta\Sigma^n}\sigma^n_{,\alpha}(\varphi,\psi)
 \bigg) \!
 \bigg]
 \sigma^m_{,i}(\varphi,\psi)
 \frac{\delta_l}{\delta\Sigma^m}
 \frac{\delta\Gamma}{\delta\psi^*_\alpha}\nonumber\\
 &&
 -
 (-1)^{\varepsilon_\alpha\varepsilon_m}\sigma^m_{,\alpha}(\varphi,\psi)
 \frac{\delta_l}{\delta\Sigma^m}
 \frac{\delta\Gamma}{\delta\psi^*_\alpha}\bigg\}.
\end{eqnarray}
 In eq.~(13), we have assumed the notation
\[
 \frac{\delta\Gamma}{\delta\phi^A}\equiv
 \bigg(
 \frac{\delta\Gamma}{\delta\varphi^i},\,
 \frac{\delta\Gamma}{\delta\psi^\alpha}
 \bigg)
\]
 and introduced the operator $\hat{\varphi^i}$
\[
 \hat{\varphi^i}=\varphi^i+i\hbar(G^{''-1})^{i\sigma}
 \frac{\delta_l}{\delta\Phi^\sigma},Œ\]
 where
\[
 \frac{\delta_l N_\sigma}{\delta\Phi^\rho}\equiv-(G^{''})_{\rho\sigma},
 \;\;\;\;
 (G^{''-1})^{\rho\delta}(G^{''})_{\delta\sigma}=\delta^\rho_\sigma\,,
\]
\begin{eqnarray*}
 \Phi^\sigma=(\varphi^i,\Sigma^m),\;\;\;
 N_\sigma=
 \bigg(-\frac{\delta\Gamma}{\delta\varphi^i}+
 \frac{\delta\Gamma}{\delta\Sigma^m}\sigma^m_{,i}(\varphi,\psi),
 -\frac{\delta\Gamma}{\delta\Sigma^m}\bigg).
\end{eqnarray*}

 The reader may profit by considering the above results in the particular
 cases of theories where either composite or external fields alone are
 present.

 Let us first turn to a quantum theory with composite fields only,
 which corresponds to the generating functional $Z=Z(J,L,\phi^*)$
 of Green's functions in eq.~(1).

 The Ward identities for the functional (1) can be derived from eqs.~(10),
 (11) which determine the Ward identities for the generating functional (7)
 of Green's functions with composite and external fields combined. Thus,
 we have
\[
 J_A\frac{\delta Z}{\delta\phi^*_A}
 +L_m\sigma^m_{,A}\left(\frac{\hbar}{i}
 \frac{\delta}{\delta J}\right)\frac{\delta Z}{\delta\phi^*_A}=0.
\]
 The above relation is obtained by integrating eq.~(10) over the external
 fields $\psi^\alpha$ with the weight functional
 $\exp{\left\{i/\hbar{\cal Y}_\alpha\psi^\alpha\right\}}$.

 The corresponding Ward identities for the generating functional
 $W=W(J,L,\phi^*)$ of connected Green's functions are consequently
 given by
\[
 J_A\frac{\delta W}{\delta\phi^*_A}+L_m\sigma^m_{,A}
 \left(\frac{\delta W}{\delta J}+
 \frac{\hbar}{i}\frac{\delta}{\delta J}\right)
 \frac{\delta W}{\delta\phi^*_A}=0,
\]
 which, in turn, implies the following Ward identities for the generating
 functional $\Gamma=\Gamma(\phi,\Sigma,\phi^*)$ of vertex functions:
\[
 \frac{1}{2}(\Gamma,\Gamma)+\frac{\delta\Gamma}
 {\delta\Sigma^m}\Biggl(\sigma^m_{,A}({\hat \phi})-\sigma^m_{,A}(\phi)Π\Biggr)\frac{\delta\Gamma}{\delta\phi^{*}_{A}}=0.
\]
 Here, $\hat{\phi}^A$ stand for the operators
\[
 \hat{\phi}^A=\phi^A+i\hbar(Q^{''-1})^{Ap}
 \frac{\delta_l}{\delta F^p},
\]
 defined as
\[
 \frac{\delta_l E_q}{\delta F^p}\equiv-(Q^{''})_{pq},
 \;\;\;\;
 (Q^{''-1})^{pr}(Q^{''})_{rq}=\delta^p_q\,,
\]
\begin{eqnarray*}
 F^p=(\phi^A,\Sigma^m),\;\;\;
 E_p=
 \bigg(-\frac{\delta\Gamma}{\delta\phi^A}+
 \frac{\delta\Gamma}{\delta\Sigma^m}\sigma^m_{,A}(\phi),
 -\frac{\delta\Gamma}{\delta\Sigma^m}\bigg).
\end{eqnarray*}

 Note that the above Ward identities for a quantum theory with composite
 fields coincide with the results obtained in Ref.~\cite{LO}.

 We next consider the case of a quantum theory with external fields only,
 evidently corresponding to the assumptions $\sigma^m(\phi)=0$, $L=0$.
 Within these restrictions, the generating functional (7) of Green's
 functions, obviously, reduces to ${\cal Z}={\cal Z}({\cal J},\psi,\phi^*)$,
 while its Ward identities, determined by eqs.~(10), (11), accordingly take
 on the form
\[
 \left(i\hbar\Delta_\psi+{\cal J}_i\frac{\delta}
 {\delta\varphi^{*}_i}\right){\cal Z}=0.
\]
 This implies the following Ward identities for the generating functional
 ${\cal W}={\cal W}({\cal J},\psi,\phi^*)$ of connected Green's functions:
\[
 \left(i\hbar\Delta_\psi+{\cal J}_i\frac{\delta}
 {\delta\varphi^{*}_i}\right){\cal W}=
 \frac{\delta{\cal W}}{\delta\psi^\alpha}
 \frac{\delta{\cal W}}{\delta\psi^*_\alpha}\,.
\]
 Finally, one readily establishes the fact that the Ward identities
 for the corresponding generating functional
 $\Gamma=\Gamma(\varphi,\psi,\phi^*)$ of vertex functions can be
 represented as
\[
 \frac{1}{2}(\Gamma,\Gamma)=i\hbar\Delta_\psi\Gamma-
 i\hbar(\Gamma^{''-1})^{ij}
 \bigg(\frac{\delta_l}{\delta\varphi^j}
 \frac{\delta\Gamma}{\delta\psi^\alpha}\bigg)Π\bigg(\frac{\delta}{\delta\psi^*_\alpha}
 \frac{\delta\Gamma}{\delta\varphi^i}\bigg),
\]
 where
\[
 (\Gamma^{''-1})^{ik}(\Gamma^{''})_{kj}=\delta^i_j\,,\;\;\;
 (\Gamma^{''})_{ij}\equiv\frac{\delta_l}{\delta\varphi^i}
 \frac{\delta\Gamma}{\delta\varphi^j}\,.
\]

 The above Ward identities for a quantum theory with external fields
 coincide with the results of Ref.~\cite{ext}.

 Consider now the change of the generating functionals ${\cal Z}$,
 ${\cal W}$, $\Gamma$ with composite and external fields under a
 variation of the gauge fermion $\Psi$ chosen in the most general form
 of an operator-valued functional, i.e.
\[
 \delta\Psi\bigg(\phi^A,\,\phi^*_A;\,
 \frac{\delta_l}{\delta\phi^A},\,\frac{\delta}{\delta\phi^*_A}\bigg)=
 \delta\Psi\bigg(\varphi^i,\,\psi^\alpha,\,\phi^*_A;\,
 \frac{\delta_l}{\delta\varphi^i},\,\frac{\delta_l}{\delta\psi^\alpha},\,
 \frac{\delta}{\delta\phi^*_A}\bigg).
\]
 Clearly, the gauge enters the generating functional ${\cal Z}$
 only in the gauge-fixed quantum action $S_{\rm ext}$, eq.~(2), whose
 variation reads
\[
 \delta\bigg(\!\exp\bigg\{\frac{i}{\hbar}S_{\rm ext}\bigg\}\bigg)
 =\hat{T}(\delta X)\exp\bigg\{\frac{i}{\hbar}S_{\rm ext}\bigg\},
\]
 where $\delta X$ is related to $\delta\Psi$ through a certain linear
 operator-valued transformation. The explicit form of $\delta X$ is not
 essential for the following treatment; nevertheless it is always possible to
 choose the operator ordering in such a way that
\begin{eqnarray}
 &&
 \delta X\bigg(\varphi^i,\psi^\alpha,\phi^*_A;\,
 \frac{\delta_l}{\delta\varphi^i},\frac{\delta_l}{\delta\psi^\alpha},
 \frac{\delta}{\delta\phi^*_A}\bigg)=
 \delta X^{(0)}\bigg(\varphi^i,\psi^\alpha,\phi^*_A;\,
 \frac{\delta_l}{\delta\psi^\alpha},\frac{\delta}{\delta\phi^*_A}\bigg)
 \nonumber\\
 &&
 +
 \sum_{N=1}\frac{\delta_l}{\delta\varphi^{i_1}}\ldots
 \frac{\delta_l}{\delta\varphi^{i_N}}
 \delta X^{(i_1\ldots i_N)}\bigg(\varphi^i,\psi^\alpha,\phi^*_A;\,
 \frac{\delta_l}{\delta\psi^\alpha},\frac{\delta}{\delta\phi^*_A}\bigg).
\end{eqnarray}
 With allowance for eqs.~(5)--(7), the variation of the functionalΠ${\cal Z}({\cal J},L,\psi,\phi^*)$ reads
\begin{eqnarray}
 \delta{\cal Z}({\cal J},L,\psi,\phi^*)=
 \int d\varphi\;\exp\bigg(\frac{i}{\hbar}{\cal J}_i\varphi^i+
 L_m\sigma^m(\varphi,\psi)\bigg)\,\times\nonumber\\
 \times
 \Delta\!\bigg(\delta X\exp\bigg\{\frac{i}{\hbar}
 S_{\rm ext}(\varphi,\psi,\phi^{*})\bigg\}\bigg).
\end{eqnarray}
 Then,
 performing in eq.~(15) integration by parts and taking the relations (9),
 (11), (14) into account, we transform the variation of the functional
 ${\cal Z}({\cal J},L,\psi,\phi^*)$ into the form
\begin{eqnarray}
 \delta{\cal Z}=\frac{1}{i\hbar}\hat{\omega}\delta\tilde{X}{\cal Z},
\end{eqnarray}
 where
$$
 \delta\!\tilde{X}=\delta\! X \! \bigg(
 \frac{\hbar}{i}\frac{\delta}{\delta{\cal J}_i},\,
 \psi^\alpha\!,\,\phi^*_A; (-1)^{\varepsilon_i}\frac{1}{i\hbar}{\cal J}_i,
 \frac{\delta_l}{\delta\psi^\alpha}
 +
 \frac{1}{i\hbar}(-1)^{\varepsilon_\alpha}L_m\sigma^m_{,\alpha}
 \bigg(
 \frac{\hbar}{i}\frac{\delta}{\delta{\cal J}},\psi\bigg),\,
 \frac{\delta}{\delta\phi^*_A}\bigg).
$$

 In terms of the generating functional
 ${\cal W}({\cal J},L,\psi,\phi^*)$ eq.~(16) can be represented as
 follows:
\begin{eqnarray}
 \delta{\cal W}=-\hat{Q}\langle\delta\tilde{X}\rangle,
\end{eqnarray}
 where $\langle\delta\tilde{X}\rangle$ is the vacuum expectation of the
 operator-valued functional $\delta\tilde{X}$
\begin{eqnarray*}
 &&\hspace*{-1cm}
 \langle\delta\tilde{X}\rangle=\delta X
 \bigg(\frac{\delta{\cal W}}{\delta{\cal J}_i}+\frac{\hbar}{i}
 \frac{\delta}{\delta{\cal J}_i},\,\psi^\alpha,\,\phi^*_A\,;
 (-1)^{\varepsilon_i}\frac{1}{i\hbar}{\cal J}_i,\nonumber\\
 &&\hspace*{-1cm}
 \frac{i}{\hbar}\frac{\delta_l{\cal W}}{\delta\psi^\alpha}+
 \frac{\delta_l}{\delta\psi^\alpha}
 +
 \frac{1}{i\hbar}(-1)^{\varepsilon_\alpha}L_m\sigma^m_{,\alpha}
 \bigg(
 \frac{\delta{\cal W}}{\delta{\cal J}}
 +Π\frac{\hbar}{i}\frac{\delta}{\delta{\cal J}},\psi
 \bigg),
 \frac{i}{\hbar}\frac{\delta{\cal W}}{\delta\phi^*_A}+
 \frac{\delta}{\delta\phi^*_A}\bigg),
\end{eqnarray*}
 and $\hat{Q}$ stands for an operator given by the rule
\begin{eqnarray}
 \hat{Q}=\exp\bigg\{\!\!-\frac{i}{\hbar}{\cal W}\bigg\}\hat{\omega}
 \exp\bigg\{\frac{i}{\hbar}{\cal W}\bigg\}.
\end{eqnarray}

 From eq.~(18) it follows, in particular, that the breakdown of nilpotency
 in the case of $\hat{\omega}$ is also inherited by $\hat{Q}$, i.e.
\[
 \hat{Q}^2=i\hbar(-1)^{\varepsilon_i}L_m
 \sigma^m_{,i\alpha}
 \left(\frac{\delta W}{\delta{\cal J}}
 +
 \frac{\hbar}{i}\frac{\delta}{\delta{\cal J}},\psi\right)
 \left[
 \left(\frac{\delta}{\delta\psi^*_\alpha}
 +
 \frac{i}{\hbar}
 \frac{\delta{\cal W}}{{\delta\psi^*_\alpha}}\right) \!
 \left(\frac{\delta}{\delta\varphi^*_i}
 +
 \frac{i}{\hbar}
 \frac{\delta{\cal W}}{{\delta\varphi^*_i}}\right)
 \right].
\]

 By virtue of the Ward identities (12) for the functional
 ${\cal W}({\cal J},L,\psi,\phi^*)$, eq.~(18) admits of the representation
\begin{eqnarray}
 \hat{Q}=\hat{\Omega}-\frac{\delta{\cal W}}{\delta\psi^\alpha}
 \frac{\delta}{\delta\psi^*_\alpha}-(-1)^{\varepsilon_\alpha}
 \frac{\delta{\cal W}}{\delta\psi^*_\alpha}\frac{\delta_l}
 {\delta\psi^\alpha}\,.
\end{eqnarray}

 In order to derive the form of gauge dependence of
 $\Gamma=\Gamma(\varphi,\Sigma,\psi,\phi^*)$, we observe that
 $\delta\Gamma=\delta{\cal W}$. Hence,
\begin{eqnarray}
 \delta\Gamma=-\hat{q}
 \langle\langle\delta\tilde{X}\rangle\rangle,
\end{eqnarray}
 where  $\langle\langle\delta\tilde{X}\rangle\rangle$ and $\hat{q}$ are the
 values related through the Legendre transformation to
 $\langle\delta\tilde{X}\rangle$ and $\hat{Q}$ in eq.~(17). Namely, the
 functional $\langle\langle\delta\tilde{X}\rangle\rangle$ has the formŒ\begin{eqnarray*}
 &&\hspace*{-0.7cm}
 \langle\langle\delta\tilde{X}\rangle\rangle
 =\delta X\bigg(\hat{\varphi}^i,\;
 \psi^\alpha,\;
 \phi^*_A;\;
 \frac{i}{\hbar}(-1)^{\varepsilon_i}\bigg(
 \frac{\delta\Gamma}{\delta\varphi^i}
 -
 \frac{\delta\Gamma}{\delta\Sigma^m}\sigma^m_{,i}(\varphi,\psi)
 \bigg),\nonumber\\
 &&\hspace*{-0.7cm}
 \frac{\delta_l}{\delta\psi^\alpha}
 +
 \frac{i}{\hbar}(-1)^{\varepsilon_\alpha}\bigg(
 \frac{\delta\Gamma}{\delta\psi^\alpha}
 -
 \frac{\delta\Gamma}{\delta\Sigma^m}\sigma^m_{,\alpha}(\varphi,\psi)
 \bigg)\nonumber\\
 &&\hspace*{-0.7cm}
 -(-1)^{\varepsilon_\alpha(\varepsilon_\rho+1)}
 (G^{''-1})^{\rho\sigma}
 \bigg[
 \frac{\delta_l}{\delta\Phi^\sigma}
 \bigg(
 \frac{\delta\Gamma}{\delta\psi^\alpha}
 -
 \frac{\delta\Gamma}{\delta\Sigma^m}\sigma^m_{,\alpha}(\varphi,\psi)
 \bigg)
 \bigg]
 \frac{\delta_l}{\delta\Phi^\rho}\nonumber\\
 &&\hspace*{-0.7cm}
 -
 (-1)^{\varepsilon_\alpha(\varepsilon_m+1)}
 \sigma^m_{,\alpha}(\varphi,\psi)\frac{\delta_l}{\delta\Sigma^m}
 +
 \frac{i}{\hbar}(-1)^{\varepsilon_\alpha}
 \frac{\delta\Gamma}{\delta\Sigma^m}
 \sigma^m_{,\alpha}(\hat{\varphi},\psi)\nonumber\\
 &&\hspace*{-0.7cm}
 +(-1)^{\varepsilon_\alpha+
 \varepsilon_i(\varepsilon_\alpha+\varepsilon_m+1)}
 (G^{''-1})^{i\sigma}
 \bigg[
 \frac{\delta_l}{\delta\Phi^\sigma}
 \bigg(
 \frac{\delta\Gamma}{\delta\psi^\alpha}
 -
 \frac{\delta\Gamma}{\delta\Sigma^n}\sigma^n_{,\alpha}(\varphi,\psi)
 \bigg)
 \bigg]Π\sigma^m_{,i}(\varphi,\psi)\frac{\delta}{\delta\Sigma^m},
 \nonumber\\
 &&\hspace*{-0.7cm}
 \frac{\delta}{\delta\phi^*_A}
 +
 \frac{i}{\hbar}\frac{\delta\Gamma}{\delta\phi^*_A}
 -
 (-1)^{\varepsilon_\rho(\varepsilon_A+1)}
 (G^{''-1})^{\rho\sigma}
 \bigg(\frac{\delta_l}{\delta\Phi^\sigma}\frac{\delta\Gamma}{\delta\phi^*_A}
 \bigg)\frac{\delta_l}{\delta\Phi^\rho}
 \nonumber\\
 &&\hspace*{-0.7cm}
 +
 (-1)^{\varepsilon_i(\varepsilon_A+\varepsilon_m)}
 (G^{''-1})^{i\sigma}
 \bigg(\frac{\delta_l}{\delta\Phi^\sigma}\frac{\delta\Gamma}{\delta\phi^*_A}
 \bigg)\sigma^m_{,i}(\varphi,\psi)\frac{\delta_l}{\delta\Sigma^m}\bigg).
\end{eqnarray*}
 Meanwhile, the operator $\hat{q}$ admits of the representation
\begin{eqnarray}
 &&\hspace*{-0.7cm}
 \hat{q}=i\hbar\bigg\{
 (-1)^{\varepsilon_\alpha}
 \frac{\delta_l}{\delta\psi^\alpha}
 -
 (-1)^{\varepsilon_\alpha\varepsilon_m}
 \sigma^m_{,\alpha}(\varphi,\psi)\frac{\delta_l}{\delta\Sigma^m}
 \nonumber\\
 &&\hspace*{-0.7cm}
 -(-1)^{\varepsilon_\alpha\varepsilon_\rho}
 (G^{''-1})^{\rho\sigma}
 \bigg[
 \frac{\delta_l}{\delta\Phi^\sigma}
 \bigg(
 \frac{\delta\Gamma}{\delta\psi^\alpha}
 -
 \frac{\delta\Gamma}{\delta\Sigma^m}\sigma^m_{,\alpha}(\varphi,\psi)
 \bigg)
 \bigg]
 \frac{\delta_l}{\delta\Phi^\rho}
 \nonumber\\
 &&\hspace*{-0.7cm}
 +
 (-1)^{\varepsilon_i(\varepsilon_\alpha+\varepsilon_m+1)}
 (G^{''-1})^{i\sigma}
 \bigg[
 \frac{\delta_l}{\delta\Phi^\sigma}
 \bigg(
 \frac{\delta\Gamma}{\delta\psi^\alpha}
 -Π\frac{\delta\Gamma}{\delta\Sigma^n}\sigma^n_{,\alpha}(\varphi,\psi)
 \bigg)
 \bigg]
 \sigma^m_{,i}(\varphi,\psi)\frac{\delta_l}{\delta\Sigma^m}
 \bigg\}\times
 \nonumber\\
 &&\hspace*{-0.7cm}
 \times
 \bigg\{
 \frac{\delta}{\delta\psi^*_\alpha}
 -
 (-1)^{\varepsilon_\rho(\varepsilon_\alpha+1)}(G^{''-1})^{\rho\sigma}
 \bigg(\frac{\delta_l}{\delta\Phi^\sigma}
 \frac{\delta\Gamma}{\delta\psi^*_\alpha}\bigg)
 \frac{\delta_l}{\delta\Phi^\rho}
 \nonumber\\
 &&\hspace*{-0.7cm}
 +
 (-1)^{\varepsilon_i(\varepsilon_\alpha+\varepsilon_m)}
 (G^{''-1})^{i\sigma}
 \bigg(
 \frac{\delta_l}{\delta\Phi^\sigma}
 \frac{\delta\Gamma}{\delta\psi^*_\alpha}\bigg)
 \sigma^m_{,i}(\varphi,\psi)
 \frac{\delta_l}{\delta\Sigma^m}
 \bigg\}
 \nonumber\\
 &&\hspace*{-0.7cm}
 -
 \bigg[
 \frac{\delta\Gamma}{\delta\phi^A}
 -
 \bigg(
 \frac{\delta\Gamma}{\delta\Sigma^n}\sigma^n_{,A}(\varphi,\psi)
 -
 \frac{\delta\Gamma}{\delta\Sigma^n}\sigma^n_{,A}(\hat{\varphi},\psi)
 \bigg)
 \bigg]\times
 \nonumber\\
 &&\hspace*{-0.7cm}
 \times
 \bigg\{
 \frac{\delta}{\delta\phi^*_A}
 -(-1)^{\varepsilon_\rho(\varepsilon_A+1)}(G^{''-1})^{\rho\sigma}
 \bigg(\frac{\delta_l}{\delta\Phi^\sigma}
 \frac{\delta\Gamma}{\delta\phi^*_A}
 \bigg)
 \frac{\delta_l}{\delta\Phi^\rho}
 \nonumber\\
 &&\hspace*{-0.7cm}
 +Π(-1)^{\varepsilon_i(\varepsilon_A+\varepsilon_m)}(G^{''-1})^{i\sigma}
 \bigg(\frac{\delta_l}{\delta\Phi^\sigma}
 \frac{\delta\Gamma}{\delta\phi^*_A}
 \bigg)
 \sigma^m_{,i}(\varphi,\psi)
 \frac{\delta_l}{\delta\Sigma^m}
 \bigg\}
 \nonumber\\
 &&\hspace*{-0.7cm}
 -
 \frac{\delta\Gamma}{\delta\psi^*_\alpha}
 \bigg\{
 (-1)^{\varepsilon_\alpha}
 \frac{\delta_l}{\delta\psi^\alpha}
 -
 (-1)^{\varepsilon_\alpha\varepsilon_m}
 \sigma^m_{,\alpha}(\varphi,\psi)\frac{\delta_l}{\delta\Sigma^m}
 \nonumber\\
 &&\hspace*{-0.7cm}
 -(-1)^{\varepsilon_\alpha\varepsilon_\rho}
 (G^{''-1})^{\rho\sigma}
 \bigg[
 \frac{\delta_l}{\delta\Phi^\sigma}
 \bigg(
 \frac{\delta\Gamma}{\delta\psi^\alpha}
 -
 \frac{\delta\Gamma}{\delta\Sigma^m}\sigma^m_{,\alpha}(\varphi,\psi)
 \bigg)
 \bigg]
 \frac{\delta_l}{\delta\Phi^\rho}
 \nonumber\\
 &&\hspace*{-0.7cm}
 +
 (-1)^{\varepsilon_i(\varepsilon_\alpha+\varepsilon_m+1})
 (G^{''-1})^{i\sigma}
 \bigg[
 \frac{\delta_l}{\delta\Phi^\sigma}
 \bigg(
 \frac{\delta\Gamma}{\delta\psi^\alpha}
 -
 \frac{\delta\Gamma}{\delta\Sigma^n}\sigma^n_{,\alpha}(\varphi,\psi)
 \bigg)
 \bigg]
 \sigma^m_{,i}(\varphi,\psi)\frac{\delta_l}{\delta\Sigma^m}
 \bigg\}.\nonumber\\
\end{eqnarray}
 At the same time, the algebraic properties of $\hat{Q}$, eq.~(18),
 obviously imply the breakdown of nilpotency for the operator $\hat{q}$
 in the general case of a theory with composite and external fields.


\section{Quantum Theories with Composite and External
         Fields in the BLT Approach}

 Consider now the quantum properties of general gauge theories with
 composite and external fields in the framework of the BLT formalism
 \cite{BLT}. For this purpose we remind that the quantization rules
 \cite{BLT} imply introducing a set of fields $\phi^A$ and a set of the
 corresponding antifields $\phi^*_{Aa}$, $\bar{\phi}_A$, with
\[
 \varepsilon(\phi^A)=\varepsilon_A,\;\;
 \varepsilon(\phi^*_{Aa})=\varepsilon_A+1,\;\;
 \varepsilon(\bar{\phi}_A)=\varepsilon_A.
\]
 The doublets of antifields $\phi^*_{Aa}$ play the role of sources of
 BRST and antiBRST transformations, while the antifields $\bar{\phi}_A$ are
 the sources of mixed BRST and antiBRST transformations.
 The structure of the configuration space of the fields $\phi^A$ in the BLT
 approach is identical (for any given gauge theory) with that of the BV
 formalism. Given this, when considered in the framework of the BLT method,
 the fields are combined into irreducible, completely symmetric,
 Sp(2)-tensors \cite{BLT}.

 The extended generating functional $Z(J,L,\phi^*,\bar{\phi})$ of Green's
 functions with composite fields is constructed within the BLT
 formalism by the rule (see, for example, Ref.~\cite{LOR})
\begin{eqnarray}
 Z(J,\phi^*,\bar{\phi})&=&\int d\phi\;\exp\bigg\{\frac{i}{\hbar}\bigg(
 S_{\rm ext}(\phi,\phi^{*},\bar{\phi})+J_A\phi^A+L_m\sigma^m(\phi)
 \bigg)\bigg\},
\end{eqnarray}
 where $S_{\rm ext}=S_{\rm ext}(\phi,\phi^{*},\bar{\phi})$ is the
 gauge-fixed quantum action defined as
\begin{equation}
 \exp\bigg\{\frac{i}{\hbar}S_{\rm ext}\bigg\}=\exp\bigg(\!\!-i\hbar
 \hat{T}(F)\bigg)\exp\bigg\{\frac{i}{\hbar}S\bigg\}.
\end{equation}
 In eq.~(23), $S=S(\phi,\phi^*,\bar{\phi})$ is a bosonic functional
 satisfying the equations
\begin{eqnarray}
 \frac{1}{2}(S,S)^a+V^aS=i\hbar\Delta^aS,
\end{eqnarray}
 or equivalently
\begin{eqnarray}
 \bar{\Delta}^a\exp\bigg\{\frac{i}{\hbar}S\bigg\}=0,\;\;\;
 \bar{\Delta}^a\equiv\Delta^a+\frac{i}{\hbar}V^a,
\end{eqnarray}
 with the boundary condition
\begin{eqnarray*}
 S|_{\phi^{*}=\bar{\phi}=\hbar=0}={\cal S}Œ\end{eqnarray*}
 (again ${\cal S}$ is the classical action). At the same time,
 $\hat{T}(F)$ is an operator of the form
\begin{equation}
 \hat{T}(F)=\frac{1}{2}\varepsilon_{ab}[\bar{\Delta}^b,[\bar{\Delta}^a,
 F]_{-}]_{+}\,,
\end{equation}
 where $F$ is a bosonic (generally, operator-valued) gauge-fixing functional.

 In eqs.~(24)--(26) we use the extended antibrackets,
 defined for two arbitrary functionals $F=F(\phi,\phi^*,\bar{\phi})$,
 $G=G(\phi,\phi^*,\bar{\phi})$ by the rule \cite{BLT}
\begin{eqnarray*}
 (F,G)^a=\frac{\delta F}{\delta\phi^A}\frac{\delta G}{\delta
 \phi^{*}_{Aa}}-
 (-1)^{(\varepsilon(F)+1)(\varepsilon(G)+1)}
 \frac{\delta G}{\delta\phi^A}\frac{\delta F} {\delta\phi^{*}_{Aa}},
\end{eqnarray*}
 as well as the operators $\Delta^a$, $V^a$
\begin{eqnarray*}
 \Delta^a=(-1)^{\varepsilon_A}\frac{\delta_l}{\delta\phi^A}\frac
 {\delta}{\delta\phi^{*}_{Aa}}\;,\;\;
 V^a=\varepsilon^{ab}\phi^{*}_{Ab}\frac{\delta}{\delta\bar\phi_A}\;
\end{eqnarray*}
 with the properties
\begin{eqnarray*}
 \Delta^{\{a}\Delta^{b\}}=0,\;\;V^{\{a}V^{b\}}=0,\;\;
 \Delta^{\{a}V^{b\}}+V^{\{a}\Delta^{b\}}=0.
\end{eqnarray*}
 From the above algebraic properties, obviously, follow the identities
 $[\hat{T}(F),\,\bar{\Delta}^a]=0$, which imply that the functional
 $S_{\rm ext}$ in eq.~(23) satisfies equations of the same form as
 in (25), i.e.
\begin{eqnarray}
 \bar{\Delta}^a\exp\bigg\{\frac{i}{\hbar}S_{\rm ext}\bigg\}=0.
\end{eqnarray}

 Consider now the following representation of the generating functional
 $Z(J,L,\phi^*,\bar{\phi})$ in eq.~(22):
\begin{eqnarray*}
 Z(J,L,\phi^*,\bar{\phi})=\int d\psi\;{\cal Z}
 ({\cal J},L,\psi,\phi^*,\bar{\phi})
 \exp\bigg(\frac{i}{\hbar}{\cal Y}\psi\bigg),
\end{eqnarray*}
 where ${\cal Z}({\cal J},L,\psi,\phi^*,\bar{\phi})$ is the (extended)
 generating functional of Green's functions with composite fields on the
 background of external fields $\psi^\alpha$
\begin{eqnarray}
 {\cal Z}({\cal J},L,\psi,\phi^*,\bar{\phi})=\int d\varphi\;
 \exp\bigg\{\frac{i}
 {\hbar}\bigg(S_{\rm ext}(\varphi,\psi,\phi^{*},\bar{\phi})+Π{\cal J}\varphi+L\sigma(\varphi,\psi)\bigg)\bigg\}
\end{eqnarray}
 with
\[
 \phi^A=(\varphi^i,\;\psi^\alpha),\;\;J_A=({\cal J}_i,\;{\cal Y}_\alpha).
\]

 In order to obtain the Ward identities for a general gauge theory with
 composite and external fields in the BLT quantization formalism, we apply
 a procedure quite similar to that presented in the case of the BV approach.
 Namely, by virtue of eq.~(27) for the gauge-fixed quantum action
 $S_{\rm ext}$ (23), we have
\begin{eqnarray}
 \int d\varphi\;\exp\bigg[\frac{i}{\hbar}\bigg({\cal J}_i\varphi^i
 +L_m\sigma(\varphi,\psi)\bigg)\bigg]
 \bar{\Delta}^a\exp\bigg\{\frac{i}{\hbar}S_{\rm ext}(\varphi,\psi,\phi^{*},
 \bar{\phi})\bigg\}=0.
\end{eqnarray}
 As a result of integration by parts in eq.~(29), with allowance for
 the relations
\begin{eqnarray}
 \exp\bigg\{\frac{i}{\hbar}\bigg(
 {\cal J}_i\varphi^i+L_m\sigma^m(\varphi,\psi)
 \bigg)\bigg\}
 \bar{\Delta}^a=\nonumber\\
\end{eqnarray}
\[
 =
 \bigg(
 \bar{\Delta}^a-\frac{i}{\hbar}{\cal J}_i
 \frac{\delta}{\delta\varphi^{*}_{ia}}-
 \frac{i}{\hbar}L_m\sigma^m_{,A}(\varphi,\psi)
 \frac{\delta}{\delta\phi^{*}_{Aa}}\bigg)
 \exp\bigg\{\frac{i}{\hbar}
 \bigg({\cal J}_i\varphi^i+
 L_m\sigma^m(\varphi,\psi)
 \bigg)
 \bigg\},
\]
 we find that the Ward identities for the functional
 ${\cal Z}={\cal Z}({\cal J},L,\psi,\phi^*,\bar{\phi})$ have the form
\begin{eqnarray}
 \hat{\omega}^a{\cal Z}=0,
\end{eqnarray}
 where
\begin{equation}
 \hat{\omega}^a=i\hbar{\Delta}^a_\psi-V^a
 +
 {\cal J}_i\frac{\delta}{\delta\varphi^{*}_{ia}}
 +
 L_m\sigma^m_{,A}\bigg(\frac{\hbar}{i}Π\frac{\delta}{\delta{\cal J}},\psi\bigg)
 \frac{\delta}{\delta\phi^{*}_{Aa}}\,,
\end{equation}
\[
 {\Delta}^a_\psi\equiv(-1)^{\varepsilon_\alpha}\frac{\delta_l}
 {\delta\psi^\alpha}\frac{\delta}{\delta\psi^{*}_{\alpha a}}.
\]
 Note that the operators $\hat{\omega}^a$ satisfy the relations
\[
 \hat{\omega}^{\{a}\hat{\omega}^{b\}}=
 i\hbar(-1)^{\varepsilon_i}L_m
 \sigma^m_{,i\alpha}\left(\frac{\hbar}{i}
 \frac{\delta}{\delta{\cal J}},\psi\right)
 \frac{\delta}{\delta\psi^*_{\alpha\{a}}
 \frac{\delta}{\delta\varphi^*_{ib\}}}\;,
\]
 which imply the violation of the property
 $\hat{\omega}^{\{a}\hat{\omega}^{b\}}=0$ of generalized nilpotency.

 In terms of the generating functional
 ${\cal W}={\cal W}({\cal J},L,\psi,\phi^*,\bar{\phi})$,
\[
 {\cal Z}=\exp\bigg\{\frac{i}{\hbar}{\cal W}\bigg\},
\]
 of connected Green's functions, the Ward identities (31) are
 represented as
\begin{eqnarray}
 \hat{\Omega}^a{\cal W}=\frac{\delta{\cal W}}{\delta\psi^\alpha}
 \frac{\delta{\cal W}}{\delta\psi^*_{\alpha a}}\,,
\end{eqnarray}
\[
 \hat{\Omega}^a=
 i\hbar\Delta_\psi^a
 -V^a
 +{\cal J}_i\frac{\delta}
 {\delta\varphi^{*}_{ia}}+L_m\sigma^m_{,A}\bigg(\frac{\delta{\cal W}}
 {\delta{\cal J}}
 +\frac{\hbar}{i}\frac{\delta}{\delta{\cal J}},\psi\bigg)
 \frac{\delta}{\delta\phi^{*}_{Aa}}.
\]
 Consequently, the Ward identities for the generating functional of 1PI vertex
 functions
\[
 \Gamma(\varphi,\Sigma,\psi,\phi^*,\bar{\phi})=
 {\cal W}({\cal J},L,\psi,\phi^*,\bar{\phi})
 -
 {\cal J}_i\varphi^i
 -
 L_m\bigg(\Sigma^m+\sigma^m(\varphi,\psi)\bigg),
\]
\[Π\varphi^i=\frac{\delta{\cal W}}{\delta{\cal J}_i},\;\;\;
 \Sigma^m=\frac{\delta{\cal W}}{\delta L_m}
 -\sigma^m\bigg(\frac{\delta{\cal W}}{\delta{\cal J}},\psi\bigg)
\]
 have the form
\begin{eqnarray}
 &&\hspace*{-0.3cm}
 \frac{1}{2}(\Gamma,\Gamma)^a+V^a\Gamma+
 \frac{\delta\Gamma}{\delta\Sigma^m}
 \bigg(\sigma^m_{,A}(\hat{\varphi},\psi)
 -\sigma^m_{,A}(\varphi,\psi)\bigg)\frac{\delta\Gamma}{\delta\phi^*_{Aa}}
 =
 \nonumber\\
 &&\hspace*{-0.3cm}
 =
 i\hbar\bigg\{
 \Delta^a_\psi\Gamma-
 (-1)^{\varepsilon_\alpha\varepsilon_\rho}
 (G^{''-1})^{\rho\sigma}
 \bigg[
 \frac{\delta_l}{\delta\Phi^\sigma}
 \bigg(
 \frac{\delta\Gamma}{\delta\psi^\alpha}
 -
 \frac{\delta\Gamma}{\delta\Sigma^m}\sigma^m_{,\alpha}(\varphi,\psi)
 \bigg)
 \bigg]
 \frac{\delta_l}{\delta\Phi^\rho}
 \frac{\delta\Gamma}{\delta\psi^*_{\alpha a}}
 \nonumber\\
 &&\hspace*{-0.3cm}
 +
 (-1)^{\varepsilon_i(\varepsilon_\alpha+\varepsilon_m+1)}
 (G^{''-1})^{i\sigma}
 \bigg[
 \frac{\delta_l}{\delta\Phi^\sigma}
 \bigg(
 \frac{\delta\Gamma}{\delta\psi^\alpha}
 -
 \frac{\delta\Gamma}{\delta\Sigma^n}\sigma^n_{,\alpha}(\varphi,\psi)
 \bigg)
 \bigg]
 \sigma^m_{,i}(\varphi,\psi)
 \frac{\delta_l}{\delta\Sigma^m}
 \frac{\delta\Gamma}{\delta\psi^*_{\alpha a}}
 \nonumber\\
 &&\hspace*{-0.3cm}
 -
 (-1)^{\varepsilon_\alpha\varepsilon_m}\sigma^m_{,\alpha}(\varphi,\psi)
 \frac{\delta_l}{\delta\Sigma^m}
 \frac{\delta\Gamma}{\delta\psi^*_{\alpha a}}\bigg\}.Œ\end{eqnarray}

 As in the previous section, let us consider the particular cases
 corresponding to theories with either composite or external fields
 only.

 Namely, for a quantum theory with composite fields, considered in the
 absence of external fields, we obtain, by virtue of eq.~(31), (32),
 the following Ward identities for the generating functional
 $Z=Z(J,L,\phi^*,\bar{\phi})$ of Green's functions in eq.~(22):
\[
 J_A\frac{\delta Z}{\delta\phi^*_{Aa}}
 +
 L_m\sigma^m_{,A}\left(\frac{\hbar}{i}
 \frac{\delta}{\delta J}\right)\frac{\delta Z}
 {\delta\phi^*_{Aa}}-V^aZ=0.
\]
 Again, they can be obtained by integrating eq.~(31) over the external
 fields $\psi^\alpha$ with the weight functional
 $\exp\{i/\hbar\,{\cal Y}_\alpha\psi^\alpha\}$.

 The corresponding generating functional $W=W(J,L,\phi^*,\bar{\phi})$
 of connected Green's functions satisfies the Ward identities of the form
\[
 J_A\frac{\delta W}{\delta\phi^*_{Aa}}
 +
 L_m\sigma^a_{,A}\left(\frac{\delta W}{\delta J}
 +
 \frac{\hbar}{i}\frac{\delta}{\delta J}\right)
 \frac{\delta W}{\delta\phi^*_{Aa}}-V^aW=0.
\]
 At the same time, the Ward identities for the generating functional
 $\Gamma=\Gamma(\phi,\Sigma,\phi^*,\bar{\phi})$ of vertex functions
 are given by
\[
 \frac{1}{2}(\Gamma,\Gamma)^a+V^a\Gamma+\frac{\delta\Gamma}
 {\delta\Sigma^m}\Biggl(\sigma^m_{,A}({\hat \phi})-\sigma^m_{,A}(\phi)
 \Biggr)\frac{\delta\Gamma}{\delta\phi^{*}_{Aa}}=0.
\]

 Note that the above Ward identities for a quantum theory with composite
 fields considered in the BLT formalism coincide with the results obtained
 in Ref.~\cite{LOR}.

 Next, for a theory with external fields in the absence of composite fields
 ($\sigma^m(\phi)=0$, $L_m=0$) the generating functional (28) of Green's
 functions evidently reduces to
 ${\cal Z}={\cal Z}({\cal J},\psi,\phi^*,\bar{\phi})$, which satisfies,
 by virtue of eqs.~(31), (32), the following Ward identities:
\[
 \left(i\hbar{\Delta}^a_\psi+{\cal J}_i\frac{\delta}Π{\delta\varphi^{*}_{ia}}-V^a\right){\cal Z}=0.
\]
 Meanwhile, the Ward identities for the corresponding generating functional
 ${\cal W}={\cal W}({\cal J},\psi,\phi^*,\bar{\phi})$ of connected Green's
 functions, accordingly, take on the form
\[
 \left(i\hbar{\Delta}^a_\psi+{\cal J}_i\frac{\delta}
 {\delta\varphi^{*}_{ia}}-V^a\right){\cal W}=
 \frac{\delta{\cal W}}{\delta\psi^\alpha}
 \frac{\delta{\cal W}}{\delta\psi^*_{\alpha a}}\,.
\]
 Finally, in the case of the generating functional
 $\Gamma=\Gamma(\varphi,\psi,\phi^*,\bar{\phi})$ of vertex
 functions the Ward identities in question can be represented as
\[
 \frac{1}{2}(\Gamma,\Gamma)^a+V^a\Gamma=i\hbar\Delta_\psi^a\Gamma
 -i\hbar(\Gamma^{''-1})^{ij}
 \bigg(\frac{\delta_l}{\delta\varphi^j}
 \frac{\delta\Gamma}{\delta\psi^\alpha}\bigg)
 \bigg(\frac{\delta}{\delta\psi^*_{\alpha a}}
 \frac{\delta\Gamma}{\delta\varphi^i}\bigg).
\]

 The above Ward identities for a quantum theory with external fields
 considered in the framework of the BLT formalism coincide with the
 corresponding results of Ref.~\cite{ext}.

 Let us now study the gauge dependence of the above generating functionals
 with composite and external fields, considered in the BLT quantization
 approach, under the most general variation
$$
 \delta F\bigg(\varphi^i,\psi^\alpha,\phi^*_{Aa},\bar{\phi}_A;\,
 \frac{\delta_l}{\delta\varphi^i},\frac{\delta_l}{\delta\psi^\alpha},
 \frac{\delta}{\delta\phi^*_{Aa}},\frac{\delta}{\delta\bar{\phi}_A}\bigg)
$$
 of the gauge boson.
 From eq.~(23) it follows that
\[
 \delta\bigg(\!\exp\bigg\{\frac{i}{\hbar}S_{\rm ext}\bigg\}\bigg)
 =-i\hbar\hat{T}(\delta Y)\exp\bigg\{\frac{i}{\hbar}S_{\rm ext}\bigg\},
\]
 where $\delta Y$, related to $\delta F$ through a linear
 (operator-valued) transformation, always admits of the representation
\begin{eqnarray}
 &&
 \hspace*{-0.5cm}
 \delta  Y\bigg(\varphi^i,\psi^\alpha,\phi^*_{Aa},\bar{\phi}_A;\,
 \frac{\delta_l}{\delta\varphi^i},\frac{\delta_l}{\delta\psi^\alpha},
 \frac{\delta}{\delta\phi^*_{Aa}},\frac{\delta}{\delta\bar{\phi}_A}\bigg)=
 \nonumber\\
 &&Π\hspace*{-0.5cm}
 =\delta Y^{(0)}\bigg(\varphi^i,\psi^\alpha,\phi^*_{Aa},
 \bar{\phi}_A;\,
 \frac{\delta_l}{\delta\psi^\alpha},\frac{\delta}{\delta\phi^*_{Aa}},
 \frac{\delta}{\delta\bar{\phi}_A}\bigg)
 \nonumber\\
 &&
 \hspace*{-0.5cm}
 +
 \sum_{N=1}\frac{\delta_l}{\delta\varphi^{i_1}}\ldots
 \frac{\delta_l}{\delta\varphi^{i_N}}
 \delta Y^{(i_1\ldots i_N)}
 \bigg(\varphi^i,\psi^\alpha,\,\phi^*_{Aa},\bar{\phi}_A;\,
 \frac{\delta_l}{\delta\psi^\alpha},
 \frac{\delta}{\delta\phi^*_{Aa}},\frac{\delta}{\delta\bar{\phi}_A}\bigg).
\end{eqnarray}
 Taking eqs.~(27), (28) and (35) into account, we have
\begin{eqnarray}
 \delta{\cal Z}({\cal J},L,\psi,\phi^*,\bar{\phi})
 =\frac{i\hbar}{2}
 \varepsilon_{ab}\int d\varphi\;\exp\bigg[\frac{i}{\hbar}\bigg(
 {\cal J}_i\varphi^i+L_m\sigma^m(\varphi,\psi)\bigg)\bigg]\times
 \nonumber\\
 \times
 \bar{\Delta}^a\bar{\Delta}^b\bigg(\delta Y
 \exp\bigg\{\frac{i}{\hbar}S_{\rm ext}(\varphi,\psi,\phi^{*},\bar{\phi})
 \bigg\}\bigg),
\end{eqnarray}
 and therefore, with allowance for eqs.~(30), (32), (35), integration by
 parts in eq.~(36) yields
\begin{eqnarray}
 \delta{\cal Z}=\frac{i}{2\hbar}
 \varepsilon_{ab}\hat{\omega}^b\hat{\omega}^a\delta\tilde{Y}{\cal Z},
\end{eqnarray}
 where
\begin{eqnarray*}
 &&
 \delta\tilde{Y}=\delta Y\bigg(
 \frac{\hbar}{i}\frac{\delta}{\delta{\cal J}_i},\,
 \psi^\alpha,\,\phi^*_{Aa}\,,\bar{\phi}_{A}\,;
 (-1)^{\varepsilon_i}\frac{1}{i\hbar}{\cal J}_i,
 \\
 &&
 \frac{\delta_l}{\delta\psi^\alpha}
 +
 \frac{1}{i\hbar}(-1)^{\varepsilon_\alpha}L_m\sigma^m_{,\alpha}
 \bigg(
 \frac{\hbar}{i}\frac{\delta}{\delta{\cal J}},\psi\bigg),\,
 \frac{\delta}{\delta\phi^*_{Aa}}\,,\frac{\delta}{\delta\bar{\phi}_{A}}\bigg).
\end{eqnarray*}
 Transforming eq.~(37) in terms of the generating functional ofΠconnected Green's functions
 ${\cal W}({\cal J},L,\psi,\phi^*,\bar{\phi})$, we arrive at the
 relation
\begin{eqnarray}
 \delta{\cal W}=\frac{1}{2}\varepsilon_{ab}\hat{Q}^b\hat{Q}^a
 \langle\delta\tilde{Y}\rangle,
\end{eqnarray}
 where
\[
 \langle\delta\tilde{Y}\rangle=\delta Y
 \bigg(\frac{\delta{\cal W}}{\delta{\cal J}_i}+\frac{\hbar}{i}
 \frac{\delta}{\delta{\cal J}_i},\,\psi^\alpha,\,\phi^*_{Aa}\,,
 \bar{\phi}_A;\,
 (-1)^{\varepsilon_i}\frac{1}{i\hbar}{\cal J}_i,
\]
\[
 \frac{i}{\hbar}\frac{\delta_l{\cal W}}{\delta\psi^\alpha}+
 \frac{\delta_l}{\delta\psi^\alpha}
 +
 \frac{1}{i\hbar}(-1)^{\varepsilon_\alpha}L_m\sigma^m_{,\alpha}
 \bigg(
 \frac{\delta{\cal W}}{\delta{\cal J}}
 +
 \frac{\hbar}{i}\frac{\delta}{\delta{\cal J}},\psi
 \bigg)\!,
 \frac{i}{\hbar}\frac{\delta{\cal W}}{\delta\phi^*_{Aa}}+
 \frac{\delta}{\delta\phi^*_{Aa}},
 \frac{i}{\hbar}\frac{\delta{\cal W}}{\delta\bar{\phi}_A}+
 \frac{\delta}{\delta\bar{\phi}_A}\bigg),
\]
 while the operators $\hat{Q}^a$ are defined by
\begin{eqnarray*}
 \hat{Q}^a=\exp\bigg\{\!\!-\frac{i}{\hbar}{\cal W}\bigg\}\hat{\omega}^a
 \exp\bigg\{\frac{i}{\hbar}{\cal W}\bigg\}
\end{eqnarray*}
 and admit of the representation
\begin{eqnarray}
 \hat{Q}^a=\hat{\Omega}^a-\frac{\delta{\cal W}}{\delta\psi^\alpha}
 \frac{\delta}{\delta\psi^*_{\alpha a}}-(-1)^{\varepsilon_\alpha}
 \frac{\delta{\cal W}}{\delta\psi^*_{\alpha a}}\frac{\delta_l}
 {\delta\psi^\alpha}\,.
\end{eqnarray}
 Note that the operators $\hat{Q}^a$ satisfy the relations
\begin{eqnarray*}
 \hat{Q}^{\{a}\hat{Q}^{b\}}
 &=&i\hbar(-1)^{\varepsilon_i}L_m
 \sigma^m_{,i\alpha}
 \left(\frac{\delta W}{\delta{\cal J}}
 +
 \frac{\hbar}{i}
 \frac{\delta}{\delta{\cal J}},\psi\right)\times\\Π&&\times
 \left[
 \left(\frac{\delta}{\delta\psi^*_{\alpha\{a}}
 +
 \frac{i}{\hbar}
 \frac{\delta{\cal W}}{{\delta\psi^*_{\alpha\{a}}}\right)
 \left(\frac{\delta}{\delta\varphi^*_{ib\}}}
 +
 \frac{i}{\hbar}
 \frac{\delta{\cal W}}{\delta\varphi^*_{ib\}}}\right)
 \right],
\end{eqnarray*}
 which follow from the properties of $\hat{\omega}^a$ and imply,
 in particular, the breakdown of generalized nilpotency also in the case
 of $\hat{Q}^a$.

 Finally, with allowance for eq.~(38), the variation of the generating
 functional $\Gamma=\Gamma(\varphi,\Sigma,\psi,\phi^*,\bar{\phi})$
 of vertex Green's functions is given by
\begin{eqnarray}
 \delta\Gamma=\frac{1}{2}\varepsilon_{ab}\hat{q}^b\hat{q}^a
 \langle\langle\delta\tilde{Y}\rangle\rangle.
\end{eqnarray}
 The functional $\langle\langle\delta\tilde{Y}\rangle\rangle$ has the
 form
\begin{eqnarray*}
 &&\hspace*{-0.7cm}
 \langle\langle\delta\tilde{Y}\rangle\rangle
 =\delta Y\bigg(\hat{\varphi}^i,\;
 \psi^\alpha,\;\phi^*_{Aa},\;\bar{\phi}_A;\;
 \frac{i}{\hbar}(-1)^{\varepsilon_i}\bigg(
 \frac{\delta\Gamma}{\delta\varphi^i}
 -
 \frac{\delta\Gamma}{\delta\Sigma^m}\sigma^m_{,i}(\varphi,\psi)
 \bigg),
 \\
 &&\hspace*{-0.7cm}
 \frac{\delta_l}{\delta\psi^\alpha}
 +
 \frac{i}{\hbar}(-1)^{\varepsilon_\alpha}\bigg(
 \frac{\delta\Gamma}{\delta\psi^\alpha}
 -
 \frac{\delta\Gamma}{\delta\Sigma^m}\sigma^m_{,\alpha}(\varphi,\psi)
 \bigg)
 \\
 &&\hspace*{-0.7cm}
 -(-1)^{\varepsilon_\alpha(\varepsilon_\rho+1)}
 (G^{''-1})^{\rho\sigma}
 \bigg[
 \frac{\delta_l}{\delta\Phi^\sigma}
 \bigg(Π\frac{\delta\Gamma}{\delta\psi^\alpha}
 -
 \frac{\delta\Gamma}{\delta\Sigma^m}\sigma^m_{,\alpha}(\varphi,\psi)
 \bigg)
 \bigg]
 \frac{\delta_l}{\delta\Phi^\rho}
 \\
 &&\hspace*{-0.7cm}
 -
 (-1)^{\varepsilon_\alpha(\varepsilon_m+1)}
 \sigma^m_{,\alpha}(\varphi,\psi)\frac{\delta_l}{\delta\Sigma^m}
 +
 \frac{i}{\hbar}(-1)^{\varepsilon_\alpha}
 \frac{\delta\Gamma}{\delta\Sigma^m}
 \sigma^m_{,\alpha}(\hat{\varphi},\psi)
 \\
 &&\hspace*{-0.7cm}
 +(-1)^{\varepsilon_\alpha+\varepsilon_i(\varepsilon_\alpha+
 \varepsilon_m+1)}
 (G^{''-1})^{i\sigma}
 \bigg[
 \frac{\delta_l}{\delta\Phi^\sigma}
 \bigg(
 \frac{\delta\Gamma}{\delta\psi^\alpha}
 -
 \frac{\delta\Gamma}{\delta\Sigma^n}\sigma^n_{,\alpha}(\varphi,\psi)
 \bigg)
 \bigg]
 \sigma^m_{,i}(\varphi,\psi)\frac{\delta}{\delta\Sigma^m},
 \\
 &&\hspace*{-0.7cm}
 \frac{\delta}{\delta\phi^*_{Aa}}
 +
 \frac{i}{\hbar}\frac{\delta\Gamma}{\delta\phi^*_{Aa}}
 -
 (-1)^{\varepsilon_\rho(\varepsilon_A+1)}
 (G^{''-1})^{\rho\sigma}
 \bigg(\frac{\delta_l}{\delta\Phi^\sigma}\frac{\delta\Gamma}
 {\delta\phi^*_{Aa}}\bigg)\frac{\delta_l}{\delta\Phi^\rho}
 \\
 &&\hspace*{-0.7cm}
 +
 (-1)^{\varepsilon_i(\varepsilon_A+\varepsilon_m)}
 (G^{''-1})^{i\sigma}
 \bigg(\frac{\delta_l}{\delta\Phi^\sigma}\frac{\delta\Gamma}
 {\delta\phi^*_{Aa}}\bigg)\sigma^m_{,i}(\varphi,\psi)\frac{\delta_l}
 {\delta\Sigma^m},
 \\
 &&\hspace*{-0.7cm}
 \frac{\delta}{\delta\bar{\phi}_A}
 +Π\frac{i}{\hbar}\frac{\delta\Gamma}{\delta\bar{\phi}_A}
 -
 (-1)^{\varepsilon_\rho\varepsilon_A}
 (G^{''-1})^{\rho\sigma}
 \bigg(\frac{\delta_l}{\delta\Phi^\sigma}
 \frac{\delta\Gamma}{\delta\bar{\phi}_A}
 \bigg)\frac{\delta_l}{\delta\Phi^\rho}
 \\
 &&\hspace*{-0.7cm}
 +
 (-1)^{\varepsilon_i(\varepsilon_A+\varepsilon_m+1)}
 (G^{''-1})^{i\sigma}
 \bigg(\frac{\delta_l}{\delta\Phi^\sigma}
 \frac{\delta\Gamma}{\delta\bar{\phi}_A}
 \bigg)\sigma^m_{,i}(\varphi,\psi)
 \frac{\delta_l}{\delta\Sigma^m}\bigg).
\end{eqnarray*}
 At the same time, the operators $\hat{q}^a$ can be represented as
\begin{eqnarray}
 &&\hspace*{-0.5cm}
 \hat{q}^a=i\hbar\bigg\{
 (-1)^{\varepsilon_\alpha}
 \frac{\delta_l}{\delta\psi^\alpha}
 -
 (-1)^{\varepsilon_\alpha\varepsilon_m}
 \sigma^m_{,\alpha}(\varphi,\psi)\frac{\delta_l}{\delta\Sigma^m}
 \nonumber\\
 &&\hspace*{-0.5cm}
 -(-1)^{\varepsilon_\alpha\varepsilon_\rho}
 (G^{''-1})^{\rho\sigma}
 \bigg[
 \frac{\delta_l}{\delta\Phi^\sigma}
 \bigg(
 \frac{\delta\Gamma}{\delta\psi^\alpha}
 -
 \frac{\delta\Gamma}{\delta\Sigma^m}\sigma^m_{,\alpha}(\varphi,\psi)
 \bigg)
 \bigg]
 \frac{\delta_l}{\delta\Phi^\rho}
 \nonumber\\
 &&\hspace*{-0.5cm}
 +
 (-1)^{\varepsilon_i(\varepsilon_\alpha+\varepsilon_m+1)}
 (G^{''-1})^{i\sigma}
 \bigg[
 \frac{\delta_l}{\delta\Phi^\sigma}
 \bigg(
 \frac{\delta\Gamma}{\delta\psi^\alpha}
 -
 \frac{\delta\Gamma}{\delta\Sigma^n}\sigma^n_{,\alpha}(\varphi,\psi)
 \bigg)Π\bigg]
 \sigma^m_{,i}(\varphi,\psi)\frac{\delta_l}{\delta\Sigma^m}
 \bigg\}\times
 \nonumber\\
 &&\hspace*{-0.5cm}
 \times
 \bigg\{
 \frac{\delta}{\delta\psi^*_{\alpha a}}
 -
 (-1)^{\varepsilon_\rho(\varepsilon_\alpha+1)}(G^{''-1})^{\rho\sigma}
 \bigg(\frac{\delta_l}{\delta\Phi^\sigma}
 \frac{\delta\Gamma}{\delta\psi^*_{\alpha a}}\bigg)
 \frac{\delta_l}{\delta\Phi^\rho}
 \nonumber\\
 &&\hspace*{-0.5cm}
 +
 (-1)^{\varepsilon_i(\varepsilon_\alpha+\varepsilon_m)}
 (G^{''-1})^{i\sigma}
 \bigg(
 \frac{\delta_l}{\delta\Phi^\sigma}
 \frac{\delta\Gamma}{\delta\psi^*_{\alpha a}}\bigg)
 \sigma^m_{,i}(\varphi,\psi)
 \frac{\delta_l}{\delta\Sigma^m}
 \bigg\}
 \nonumber\\
 &&\hspace*{-0.5cm}
 -
 \bigg[
 \frac{\delta\Gamma}{\delta\phi^A}
 -
 \bigg(
 \frac{\delta\Gamma}{\delta\Sigma^n}\sigma^n_{,A}(\varphi,\psi)
 -
 \frac{\delta\Gamma}{\delta\Sigma^n}\sigma^n_{,A}(\hat{\varphi},\psi)
 \bigg)
 \bigg]\times
 \nonumber\\
 &&\hspace*{-0.5cm}
 \times
 \bigg\{
 \frac{\delta}{\delta\phi^*_{Aa}}
 -(-1)^{\varepsilon_\rho(\varepsilon_A+1)}(G^{''-1})^{\rho\sigma}
 \bigg(\frac{\delta_l}{\delta\Phi^\sigma}
 \frac{\delta\Gamma}{\delta\phi^*_{Aa}}
 \bigg)
 \frac{\delta_l}{\delta\Phi^\rho}
 \nonumber\\
 &&\hspace*{-0.5cm}
 +
 (-1)^{\varepsilon_i(\varepsilon_A+\varepsilon_m)}(G^{''-1})^{i\sigma}
 \bigg(\frac{\delta_l}{\delta\Phi^\sigma}Π\frac{\delta\Gamma}{\delta\phi^*_{Aa}}
 \bigg)
 \sigma^m_{,i}(\varphi,\psi)
 \frac{\delta_l}{\delta\Sigma^m}
 \bigg\}
 \nonumber\\
 &&\hspace*{-0.5cm}
 -
 \frac{\delta\Gamma}{\delta\psi^*_{\alpha a}}
 \bigg\{
 (-1)^{\varepsilon_\alpha}
 \frac{\delta_l}{\delta\psi^\alpha}
 -
 (-1)^{\varepsilon_\alpha\varepsilon_m}
 \sigma^m_{,\alpha}(\varphi,\psi)\frac{\delta_l}{\delta\Sigma^m}
 \nonumber\\
 &&\hspace*{-0.5cm}
 -(-1)^{\varepsilon_\alpha\varepsilon_\rho}
 (G^{''-1})^{\rho\sigma}
 \bigg[
 \frac{\delta_l}{\delta\Phi^\sigma}
 \bigg(
 \frac{\delta\Gamma}{\delta\psi^\alpha}
 -
 \frac{\delta\Gamma}{\delta\Sigma^m}\sigma^m_{,\alpha}(\varphi,\psi)
 \bigg)
 \bigg]
 \frac{\delta_l}{\delta\Phi^\rho}
 \nonumber\\
 &&\hspace*{-0.5cm}
 +
 (-1)^{\varepsilon_i(\varepsilon_\alpha+\varepsilon_m+1)}
 (G^{''-1})^{i\sigma}
 \bigg[
 \frac{\delta_l}{\delta\Phi^\sigma}
 \bigg(
 \frac{\delta\Gamma}{\delta\psi^\alpha}
 -
 \frac{\delta\Gamma}{\delta\Sigma^n}\sigma^n_{,\alpha}(\varphi,\psi)
 \bigg)
 \bigg]
 \sigma^m_{,i}(\varphi,\psi)\frac{\delta_l}{\delta\Sigma^m}
 \bigg\}
 \nonumber\\
 &&\hspace*{-0.5cm}
 -\varepsilon^{ab}\phi^*_{Aa}\bigg\{
 \frac{\delta}{\delta\bar{\phi}_A}
 -
 (-1)^{\varepsilon_\rho\varepsilon_A}
 (G^{''-1})^{\rho\sigma}
 \bigg(\frac{\delta_l}{\delta\Phi^\sigma}Π\frac{\delta\Gamma}{\delta\bar{\phi}_A}
 \bigg)\frac{\delta_l}{\delta\Phi^\rho}
 \nonumber\\
 &&\hspace*{-0.5cm}
 +
 (-1)^{\varepsilon_i(\varepsilon_A+\varepsilon_m+1)}
 (G^{''-1})^{i\sigma}
 \bigg(\frac{\delta_l}{\delta\Phi^\sigma}
 \frac{\delta\Gamma}{\delta\bar{\phi}_A}
 \bigg)\sigma^m_{,i}(\varphi,\psi)
 \frac{\delta_l}{\delta\Sigma^m}\bigg\}.
\end{eqnarray}
 Clearly, the above values $\hat{q}^a$ and
 $\langle\langle\delta\tilde{Y}\rangle\rangle$ are the Legendre transforms
 of the corresponding values $\hat{Q}^a$ and
 $\langle\delta\tilde{Y}\rangle$ in eq.~(38). As far as the operators
 $\hat{q}^a$ are concerned, this implies, in particular, the violation of
 their generalized nilpotency in the general case of a theory with composite
 and external fields.

\section{Conclusion}
 In this paper we have presented an extension of the studies
 of arbitrary quantum gauge theories with composite
 \cite{LO,LOR} and external \cite{ext} fields to the case of theories
 with composite and external fields combined.
 Namely, we considered the generating functionals of Green's functions
 with composite and external fields in the framework of the BV \cite{BV}
 and BLT \cite{BLT} quantization methods for general gauge theories. For
 these functionals we obtained, using the technique developed in
 refs.~\cite{LOR,ext}, the corresponding Ward identities, eqs.~(10),
 (12), (13), (31), (33), (34), and derived the explicit dependence on the
 most general form of gauge-fixing, eqs.~(16), (17), (20), (37), (38), (40).
 The gauge dependence is described with the help of fermionic operators (11),
 (19), (21) in the BV method, as well as with the help of doublets of
 fermionic operators (32), (39), (41) in the framework of the BLT formalism,
 which bears remarkable similarity to the cases where only composite
 \cite{LOR} or external \cite{ext} fields were present.

 At the same time, it should be noted that the most general scheme
 combining composite and external fields proves to present some essentially
 new features, which set it apart from the previous cases \cite{LO,LOR,ext}.
 First of all, we refer to the fact that the form of the Ward identities and
 gauge dependence admits, in the case under consideration, an entirely
 different character and does not follow from those derived earlier. (The
 latter, naturally, can be obtained from the general case as the
 corresponding limits.) Another notable evidence is provided by the fact
 that in the most general case of composite and external fields combined,
 the operators describing gauge dependence do not possess the algebraic
 properties of (generalized) nilpotency which hold for their counterpartsΠ\cite{LOR,ext}. The lack of generalized nilpotency has been traced back to
 expressions including $\sigma^m_{, i \alpha} (\varphi, \psi)$, i.\,e.~to
 the explicit dependence of the composite fields upon the external ones.
 Of course, this fact deserves further considerations, especially with
 respect to the effect it may have on the unitarity of the theory.
 This, however, has to be postponed to another paper.

 Let us finally remark that --- despite being much more complicated ---
 an extension of these considerations to the manifest osp(1,2)-covariant
 quantization \cite{GLM} would be desirable.


\section*{Acknowledgments}
 The authors would like to thank P.M. Lavrov, S.D. Odintsov, D.~M\"{u}lsch
 and A.A.~Reshetnyak for useful discussions. One of the authors (PM) is
 grateful for kind hospitality extended to him by the Graduate College
 "Quantum Field Theory" at the Center of Theoretical Sciences (NTZ) of
 Leipzig University during the winter half, 1997 -- 1998.

 The work has been partially supported by the Russian Foundation for
 Basic Research (RFBR), project 96--02--16017.


\end{document}